\documentclass[final,1p,times]{elsarticle} 
\usepackage{graphicx} 
\usepackage{amssymb} 
\usepackage{amsthm} 
\usepackage{lineno} 

\journal{Nuclear Physics A} 
\begin{document} 

\begin{frontmatter} 

\title{Universal Flow in the First fm/c at RHIC}

\author{\b{Joshua Vredevoogd} and Scott Pratt}

\address[a]{Department of Physics and Astronomy,
Michigan State University\\
East Lansing, Michigan 48824}

\begin{abstract} 
We will show how both elliptic and radial flßow generated during the fiÞrst fm/c at RHIC is inde- 
pendent of the state of matter and depends only on the initial energy density profiÞle. Descriptions 
based on partons or classical fiÞelds, thermalized or highly anisotropic, all lead to the same collec- 
tive velocity given a few easily satisfiÞed conditions. This signifiÞcantly narrows the uncertainty for 
initializing hydrodynamic prescriptions.
\end{abstract} 
\end{frontmatter} 

\section{Introduction and Definitions}
Colliding heavy ions at the energies available at the Relativistic Heavy Ion Collider have produced experimental signatures strongly consistent with almost fully thermalized matter.
For instance, heavier particles show a larger average transverse energy\cite{mass_order}.
In addition, the matter shows strong anisotropic collcective flow on the order of ten percent\cite{strong_flow}.
Each of these observations are consistent with early predictions from ideal hydrodynamics\cite{early_hydro}.
Two particle correlations give insight into the spatial dimensions of the source \cite{femto_over}.
These data provide minimal constraints in isolation, but in conjunction, there appear strong features.
Ideal hydrodynamics were quickly shown to describe both the strong anisotropic flow and the relatively small longitudinal size \cite{HK_hydro} 
despite the large velocity gradients required by boost invariance at central rapidity \cite{Bjork}.
However, such models required an extremely rapid thermalization of the matter and under-predicted the explosiveness of the source.
This apparent tension was referred to as the 'HBT puzzle.'

Hydrodynamics, for obvious reasons, is not capable of describing the nuclear matter immediately following the first interactions as the nuclei pass through one another.
An early model is required, but needs only to provide a description of roughly the first one fm/c, while hydrodynamics - maybe viscous hydrodynamics -  will run until $\sim10$ fm/c.
It is important to note however, that early collective, transverse acceleration is preferentially important as compared to later acceleration.
Therefore, there is a certain model dependence in the initial conditions chosen to instantiate a hydrodynamic simulation with several prospects \cite{CGC},\cite{Partons}.
Each of these models can be thought of in terms of a conserved stress-energy tensor, $ \partial^\alpha T_{\alpha \beta} = 0$, which should hold in general while the contents of the tensor will vary from model to model.
These differences can be parameterized by $\kappa$, which we define to be
\begin{equation}
T_{xx}=T_{yy} \equiv \kappa T_{00}
\end{equation}
where $\kappa$ would generally be thought of as a transverse stiffness that would vary from one-third for relativistic hydrodynamics to one-half for longitudinally free-streaming particles to unity for longitudinal coherent fields.
The result of this paper will be that the transverse flow generated at early times, as defined by
\begin{equation}
{\rm \displaystyle F_i}\equiv\frac{T_{0i}}{T_{00}},
\end{equation}
is independent of $\kappa$.
\section{Results}
Calculating the evolution of flow at early times requires only the application of the conservation equation \cite{jav}.
We impose a few simple assumptions: 
\begin{enumerate}
\item That the stress energy tensor be traceless. 
This assumption is completely valid for non-interacting particles or fields as well as for ideal hydrodynamics.
At the large energy densities during the first fm/c, this should be valid at the 10\% level.
While the transverse pressures are set by $\kappa$, this condition specifies that $T_{zz} =(1-2\kappa)T_{00}$.
\item That the longitudinal dynamics be entirely boost-invariant.
For $\eta \leq ~1$ this has been proved to be valid at roughly the 10\% level as well.
This fixes the longitudinal velocity to be $u_z = z/\tau$.
\item That the anisotropy in the stress energy tensor by purely time dependent.
This is akin to the same description of the system applying at a given time.
\end{enumerate}

Intuitively, one might expect that an increase in the value of $\kappa$ should lead directly to an increase in the transverse velocity.
For a one-dimensional system with no longitudinal expansion, the conservation of momentum would require that
\begin{equation}
\partial_\tau T_{0x} = - \partial_x T_{xx} = - \kappa \partial_x T_{00}
\end{equation}
This indicates a linear increase in the transverse flow as one increases the anisotropy.
However, the longitudinal expansion adds an additional term since 
\begin{equation}
\partial_z T_{xz} = \partial_z T_{0x} z/\tau = T_{0x}/\tau
\end{equation}
for small velocities.
Using the fact that the energy density falls according to
\begin{equation}
\partial_\tau T_{00} = - \frac{2}{\tau} (1-\kappa) T_{00}
\end{equation}
Calculating the time derivative of the flow is then straightforward
\begin{eqnarray}
\partial_\tau\frac{T_{0x}}{T_{00}}&=&\frac{\partial_\tau T_{0x}}{T_{00}}
-\frac{T_{0x}}{T_{00}}\frac{\partial_\tau T_{00}}{T_{00}}\\
\nonumber
&=&-\frac{\kappa\partial_x T_{00}+T_{0x}/\tau}{T_{00}}
+\frac{(2-2\kappa)}{\tau}\frac{T_{0x}}{T_{00}}.
\end{eqnarray}
which lead directly to
\begin{equation} \label{univ_eq}
\frac{T_{0x}}{T_{00}} = \frac{ -\partial_x T_{00}}{T_{00}} \tau
\end{equation}
This equation is independent of $\kappa$, meaning that it does not depend on the exact description of the system's dynamics beyond those stipulated in the assumptions.

\begin{figure}[ht]
\centering
\includegraphics[scale=0.25]{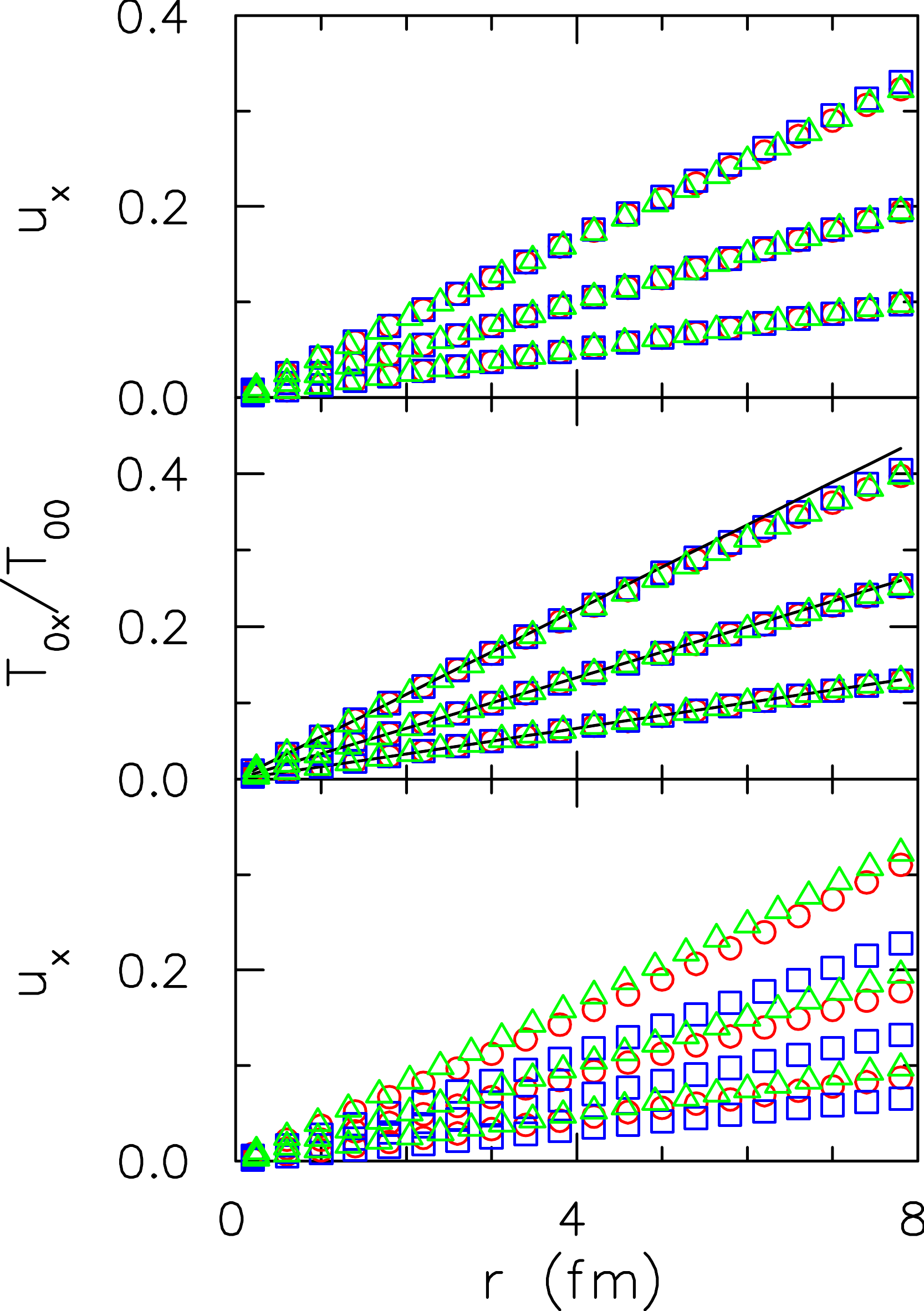}	       
\caption[]{ 
Triangles - ideal hydrodynamics ($\kappa=1/3$).  
Circles - free-streaming particles ($\kappa=1/2$).  
Squares - coherent fields ($\kappa=1$).
Each panel contains integration results from $\tau=0.3,0.6,1.0$fm/c.
Bottom panel shows the transverse velocity as defined in each model, showing vast differences between the models.
Middle panel shows the transverse flow in each model.
Top panel shows the transverse velocity as defined in ideal hydrodynamics after a rapid thermalization from each model.
(Color Online)
}
\label{univ_pdf}
\end{figure}

As shown in Figure 1, while the transverse flow of the system develops identically between the models, the collective velocities differ.
After around one fm/c, the system should thermalize.
If the transition is smooth in time, then our constraints on $\kappa$ allow for a smooth transition of anisotropy into ideal hydrodynamics.
In the case that this process occurs suddenly and simultaneously on a hypersurface parameterized by the four-vector $n_\alpha = (1,0,0,0)$, the elements $T_{\alpha0}$ are conserved.
In the spirit of the Rankine-Hugeniot equations, one can integrate the conservation equation across a shock as a model of thermalization:
\begin{equation}
0 = \int_{\tau-\delta \tau}^{\tau+\delta \tau} \left[ \partial_\tau T_{0\alpha} + \partial_i T_{i\alpha} \right] = T_{0\alpha}(\tau + \delta \tau) - T_{0\alpha}(\tau - \delta \tau)
\end{equation}
Integration across a general hypersurface would have resulted in the conservation of $n_\beta T^{\alpha \beta}$, and for a space-like surface (e.g. $n=(0,0,0,1)$)) would result in the usual Rankine-Hugeniot equations.
However, for any transition such that $n^2=+1$, $T_{0\alpha}$ is conserved exactly in some frame.
The top panel of Figure 1 shows that such a transition leads to identical velocity profiles as the initial conditions of a hydrodynamic simulation.

\begin{figure}[ht]
\centering
\includegraphics[scale=0.25]{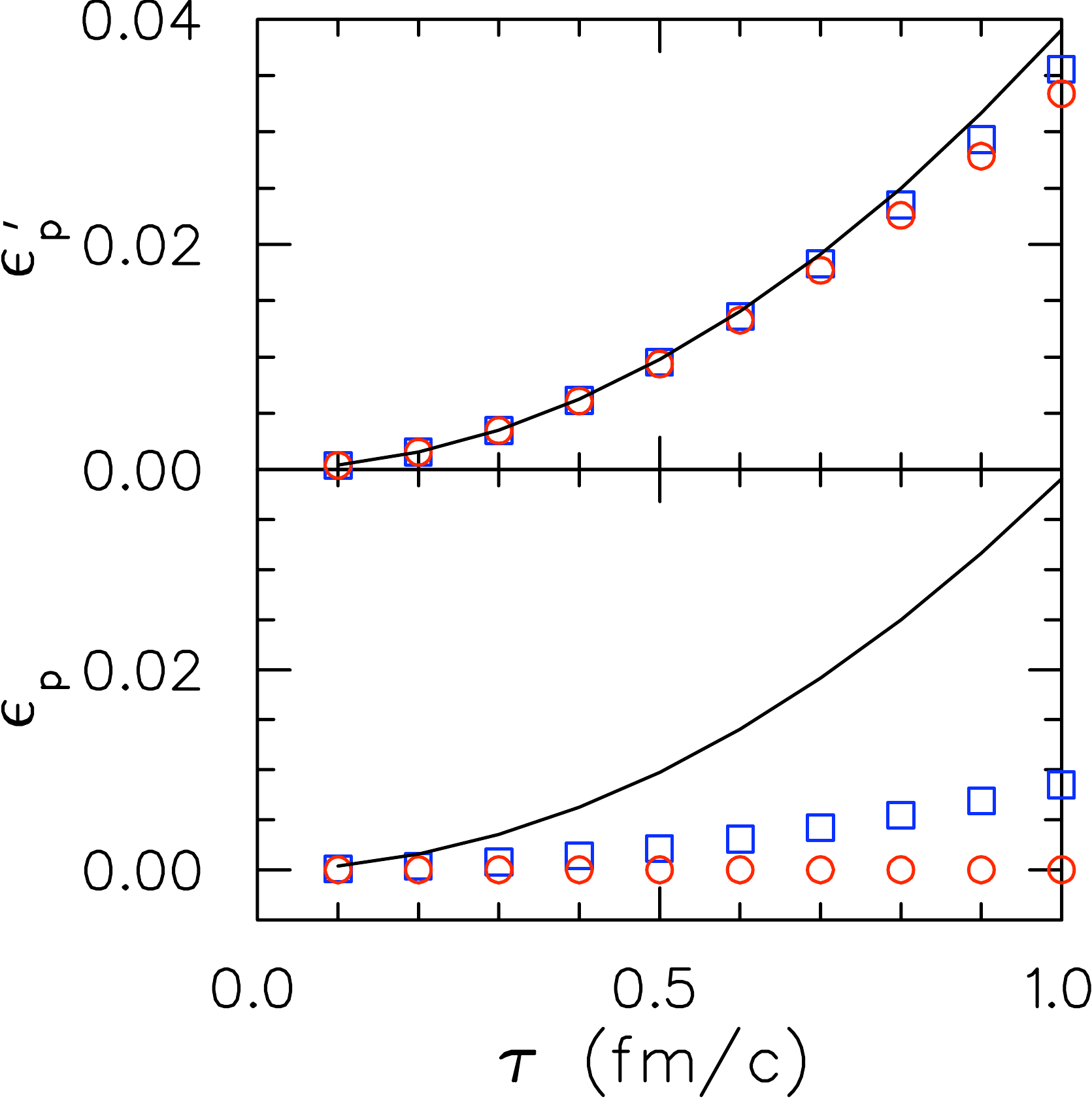}	       
\caption[]{ 
Line - ideal hydrodynamics ($\kappa=1/3$).  
Circles - free-streaming particles ($\kappa=1/2$).  
Squares - coherent fields ($\kappa=1$).
The anisotropy in momentum space as a function of $\tau$.   
Bottom panel shows that ideal hydrodynamics develops stronger anisotropic flow.
Top panel shows that rapid thermalization immediately restores the agreement of the models.
(Color Online).
}
\label{ep_pdf}
\end{figure}

As a corollary, we point out that this precisely means that anisotropic flow develops as well.
This follows clearly from eq. \ref{univ_eq} - the flow develops independent of $\kappa$ but proportional to the gradient of the energy density.
That is to say, along transverse directions in which the gradients are larger, there will still develop stronger flow.
Now, in general, the elliptic flow developed in hydrodynamics is measured using the quantity
\begin{equation}
\epsilon_p = \frac{ < T_{xx} - T_{yy}>}{ < T_{xx} + T_{yy}>}
\end{equation}
where the brackets indicate an integral over the entire plane.
In hydrodynamics, prior to freeze-out, the quantity is directly proportional to elliptic flow.
However, since our models vary in what is meant by $T_{xx}$ and $T_{yy}$, it is not clear that a change in $\epsilon_p$ (or lack thereof) results in a changed (or unchanged) elliptic flow as measured in the final state.
The bottom panel of Figure 2 shows that while hydrodynamics develops a strong signal in $\epsilon_p$, other models develop little or no signal.
But the system has yet to thermalize and the flow developed is not wasted.
The top panel of Figure 2 shows that anisotropic flow is indeed developing and is only shown after thermalization.

We have shown that early collective velocity as strong as those developed in hydrodynamics are created by any model within some mild constraints.
In addition, elliptic flow begins to develop before a hydrodynamic model can be applied.
Using initial conditions developed from principles outlined here, which we expect to be valid at the $\approx$10\% level, particle spectra and HBT radii can described at roughly the same level using a viscous hydrodynamic simulation coupled to a resonance cascade.\cite{Pratt:2008sz}

\section*{Acknowledgments} 
Support was provided by the U.S. Department of Energy, Grant No. DE-FG02-03ER41259. 

\end{document}